\documentclass[aip,amsmath,amssymb,graphics,reprint]{revtex4-1}

\usepackage[utf8]{inputenc}
\usepackage[T1]{fontenc}

\usepackage{graphicx}
\usepackage{float}
\usepackage{hyperref}
\usepackage{xcolor}

\begin{document}

\title{Paradoxical nature of negative mobility in the weak dissipation regime}

\author{Mateusz Wi\'{s}niewski}
\author{Jakub Spiechowicz}
\email[]{jakub.spiechowicz@us.edu.pl}
\affiliation{Institute of Physics, University of Silesia in Katowice, 41-500 Chorzów, Poland}

\date{\today}

\begin{abstract}
	We reinvestigate a paradigmatic model of nonequilibrium statistical physics consisting of an inertial Brownian particle in a symmetric periodic potential subjected to both a time--periodic force and a static bias. In doing so we focus on the negative mobility phenomenon in which the average velocity of the particle is opposite to the constant force acting on it. Surprisingly, we find that in the weak dissipation regime thermal fluctuations induce negative mobility much more frequently than it happens if the dissipation is stronger. In particular, for the very first time we report a parameter set in which thermal noise causes this effect in the nonlinear response regime. Moreover, we show that the coexistence of deterministic negative mobility and chaos is routinely encountered when approaching the overdamped limit in which chaos does not emerge rather than near the Hamiltonian regime of which chaos is one of the hallmarks. On the other hand, at non-zero temperature the negative mobility in the weak dissipation regime is typically affected by the weak ergodicity breaking. Our findings can be corroborated experimentally in a multitude of physical realizations including e.g.~Josephson junctions and cold atoms dwelling in optical lattices.
\end{abstract}


\maketitle

\begin{quotation}
Phenomena occurring at the border of two physical realms are often the most fascinating ones. As an example it is enough to mention the still not fully solved problem of quantum to classical transition. In this spirit we analyze a weak dissipation regime of a dynamical system in which thermal fluctuations interact with a rich complexity of deterministic dynamics. Surprisingly, the impact of weak thermal noise is not minor as it induces behavior which is present neither for a dissipationless nor an overdamped system and, counterintuitively, it happens much less frequently when the dissipation is stronger. Moreover, we demonstrate that the hallmarks of these two limiting cases may not always serve as a guide towards physical reality when approaching them by a weak or a strong dissipation regime.
\end{quotation}

\section{Introduction}
A body immersed in a fluid constantly interacts with its environment. 
For macroscopic objects we usually ignore the molecular nature of the medium and take advantage of phenomenological description valid mostly in thermal equilibrium.
In the microscale, when the body itself is a particle, the interactions with individual molecules of the environment need to be considered. They play a role of a random force which maintains the erratic motion of the system.
This leads to a model of a Brownian particle surrounded by a constantly fluctuating medium \cite{Einstein1905,Smoluchowski1906}.

When the Brownian particle is driven by external perturbation it always suffers from a friction, a systematic force proportional to its velocity. Therefore as a consequence of interactions with the environment the system experiences two effects: (i) the stochastic agitation and (ii) the dissipative force. Since they both have the same origin they are related by the celebrated \emph{fluctuation--dissipation theorem} \cite{Callen1951, Kubo1966, Landau, Marconi2008} which loosely speaking states that the response of a system to an external perturbation is determined by its fluctuation in the absence of this disturbance. The former is described by an admittance or an impedance while the latter is characterized by a correlation function of a relevant physical quantity or its fluctuation spectrum.

The fluctuation--dissipation theorem is valid for classical and quantum systems in thermal equilibrium with a perturbation applied in the linear response regime. However, there is another manifestation of this relation, also known as the \emph{second} fluctuation--dissipation theorem \cite{Kubo1966}, which survives even in nonequilibrium states. It tells that the friction is determined by a correlation of the random force originating from microscopic interactions with the environment. Equivalently, the power spectrum of thermal fluctuations is characterized by the friction. It means that the weak dissipation usually implies weak thermal noise.

A natural question arises whether weak thermal fluctuations can have significant impact on the dynamics of a system in the weak damping regime or even play a leading role in it? Thermal noise may destabilize stationary states and induce new ones which could correspond to qualitatively and quantitatively different behavior. It is the \emph{modus operandi} of effects like stochastic resonance \cite{Benzi1981,Gammaitoni1998}, noise--induced transport \cite{Reimann2002,Hanggi2009,Cubero2016} or dynamical localization \cite{Spiechowicz2017scirep,Spiechowicz2019chaos}. Thermal fluctuations acting upon a nonlinear system far from equilibrium may have particularly far--reaching consequences. This fact is rooted in two properties of such setups. Firstly, as nonlinear they are unaffected by the superposition principle and secondly, in nonequilibrium state thermodynamic laws and various symmetries such as the detailed balance generally loose their validity.

These far--reaching consequences are not rarely also counterintuitive. One of their examples is the negative mobility effect \cite{McCombie1997, Reimann1999, Cleuren2001, Eichhorn2002prl, Eichhorn2002pre, Cleuren2002, Cleuren2003, Haljas2004, Ros2005}, in which the net movement of the particle is opposite to the direction of the average force acting on it. The minimal system where this effect can be observed is a paradigmatic model of nonequilibrium statistical physics consisting of an inertial Brownian particle dwelling in a periodic potential and subjected to both a time--periodic and a static force \cite{Machura2007, Speer2007pre}. The problem of negative mobility of a Brownian particle has a long history but remains also a vibrant topic of current research \cite{Machura2007, Speer2007epl, Speer2007pre, Nagel2008, Kostur2008, Kostur2009, Hanggi2010, Eichhorn2010, Januszewski2011, Du2011, Du2012, Spiechowicz2013jstatmech, Spiechowicz2014pre, Ghosh2014, Malgaretti2014, Dandogbessi2015, Luo2016, Sarracino2016, Slapik2018, Cecconi2018, Ai2018, Cividini2018, Slapik2019prl, Slapik2019prappl, Spiechowicz2019njp, Sonker2019, Wu2019, Zhu2019, Luo2020chaos, Luo2020pre, Wisniewski2022, Wu2022, Luo2022, GR2022}. Due to the complex multidimensional parameter space of this model the vast majority of the research focused solely on selected parameter regimes. Only recently \cite{Wisniewski2022} GPU supercomputers allowed for its systematic and comprehensive exploration to draw general conclusions about the emergence of negative mobility.

In this work we revisit this system to investigate its dynamics in an unexplored limit of weak dissipation to reveal a number of paradoxes of the negative mobility phenomenon. In particular, we demonstrate that in such a situation weak thermal fluctuations has the greatest impact on the emergence of this effect. Moreover, we illustrate instances of constructive influence of thermal noise on the considered anomalous transport behavior, e.g. for the first time we report that thermal fluctuations can induce the negative mobility in the nonlinear response regime.

The paper is organized as follows.
In Section \ref{sec:model} we present the model which we use in this study and define the basic quantity of interest that we refer to throughout the text.
Next, in Section \ref{sec:methods} we describe the numerical methods that allowed us to perform the simulations and analyze the results.
Section \ref{sec:results} presents several paradoxes of negative mobility in the weak dissipation regime and examples of constructive influence of thermal fluctuations on this effect.
The last Section \ref{sec:conclusions} provides the conclusions of our work.

\section{Model} \label{sec:model}
In this paper we consider a model of a one--dimensional Brownian motion in a driven non--linear periodic system.
We assume that the Brownian particle is dwelling in a symmetric potential with a spatial period $L$, namely
\begin{equation}
	U(x) = \Delta U \sin(2\pi x/L),
\end{equation}
where $x$ is the position of the particle.
Moreover, it is driven by two external forces: time--periodic $A\cos(\Omega t)$ and constant $F$.
In addition it is exposed to thermal fluctuations modeled by $\delta$--correlated Gaussian noise, i.e.
\begin{equation}
	\langle \xi(t) \rangle = 0,\quad \langle \xi(t) \xi(s) \rangle = \delta(t-s).
\end{equation}
Such a system can be modelled by the following Langevin equation\cite{Slapik2019prl}:
\begin{equation} \label{eq:lang}
	M\ddot{x} + \Gamma\dot{x} = -\frac{\mathrm{d}U(x)}{\mathrm{d}x} + A\cos(\Omega t) + F + \sqrt{2\Gamma k_B T}\xi(t),
\end{equation}
where $M$ is the mass of the particle, $\Gamma$ represents the friction coefficient, $T$ is the temperature of the environment and dot represents differentiation with respect to the time $t$.
Thermal noise prefactor $\sqrt{2\Gamma k_B T}$ follows from the fluctuation--dissipation theorem 
\cite{Marconi2008} and ensures the Gibbs equilibrium state for vanishing external perturbations when $A = 0$ and $F = 0$.

To reduce the number of the parameters and make the model setup--independent we transform Eq.~(\ref{eq:lang}) to the dimensionless form.
Different choice of the length and the time units leads to a different form of the equation and allows to eliminate some of the parameters. 
The two most commonly used scalings rule out the mass and the friction coefficient, respectively\cite{Machura2008}.
The procedure of obtaining these scaled equations and the differences between them are discussed in Ref.~[\onlinecite{Wisniewski2022}], therefore here we will only write them down without a detailed explanation.
The first mentioned scaling results in the following equation
\begin{equation} \label{eq:lang:gam}
	\dot{v}_1 + \gamma v_1 = -\frac{\mathrm{d}\hat{U}(\hat{x})}{\mathrm{d}\hat{x}} + a\cos(\omega_1 t_1) + f + \sqrt{2\gamma D}\ \hat{\xi}(t_1),
\end{equation}
where $\hat{x}=x/L$ and $t_1=t/\tau_1$ are the scaled position and time variables and $v_1=\mathrm{d}\hat{x}/\mathrm{d}t_1$. In this case the length unit is the period of the potential $L$ and the time unit $\tau_1=L\sqrt{M/\Delta U}$ can be extracted from the equation of the frictionless motion of the particle in the periodic potential
\begin{equation}
	M\ddot{x} = -U'(x).
\end{equation}
The rescaled temperature $D=k_B T/\Delta U$ is the ratio of thermal and half of the potential barrier energies. The remaining parameters read
\begin{subequations}
\begin{align}
	&\gamma = \tau_1/\tau_0, \\ 
	&a = (L/\Delta U)A, \\
	&\omega_1 = \tau_1 \Omega, \\
	&f = (L/\Delta U) F,
\end{align}
\end{subequations}
where $\tau_0 = M/\Gamma$ stands for the so called Langevin time, i.e. the characteristic time for velocity relaxation of the Brownian particle.

The other dimensionless form of Eq.~(\ref{eq:lang}) is
\begin{equation} \label{eq:lang:mass}
	m\dot{v}_2 + v_2 = -\frac{\mathrm{d}\hat{U}(\hat{x})}{\mathrm{d}\hat{x}} + a\cos(\omega_2 t_2) + f + \sqrt{2D}\ \hat{\xi}(t_2),
\end{equation}
where $t_2$ is the scaled time  and $v_2=\mathrm{d}\hat{x}/\mathrm{d}t_2$.
Here the length unit is the same as in Eq.~(\ref{eq:lang:gam}), but the characteristic time $\tau_2=\Gamma L^2/\Delta U$ follows from the overdamped motion of the particle in the periodic potential
\begin{equation}
	\Gamma \dot{x} = -U'(x).
\end{equation}
In Eq.~(\ref{eq:lang:gam}) the mass formally scales to $m\equiv1$, but the friction coefficient $
\gamma$ remains as a parameter.
This makes this equation suitable for investigating the influence of the dissipation, especially in the limiting case of weak damping when $\gamma \to 0$.
Similarly, in Eq.~(\ref{eq:lang:mass}) the friction coefficient formally scales to $\gamma\equiv1$, but the mass $m$ can be still modified.
For this reason this scaling should be used when approaching overdamped limit, i.e. for strongly damped system with $m \to 0$.
Since the main topic of this study is the weak dissipation regime, we will stick to the first scaling and for simplicity omit the index 1 in $t_1$, $v_1$, $\omega_1$, and write $x$, $U$, $\xi$ instead of $\hat{x}$, $\hat{U}$ and $\hat{\xi}$.

There exists a multitude of physical systems which can be modeled in the framework of dynamics described by Eq.~(\ref{eq:lang:gam}) or Eq.~(\ref{eq:lang:mass}). They are superionic conductors \cite{Fulde1975,Dieterich1980}, dipoles rotating in external fields \cite{Coffey2004}, charge density waves \cite{Gruner1981}, Josephson junctions \cite{Kautz1996,Blackburn2016} and their combinations like SQUIDS \cite{Spiechowicz2015njp,Spiechowicz2015chaos} as well as cold atoms dwelling in optical lattices \cite{Lutz2013,Denisov2014}, to name only a few.
\begin{figure*}[t]
	\centering
	\includegraphics[width=\linewidth]{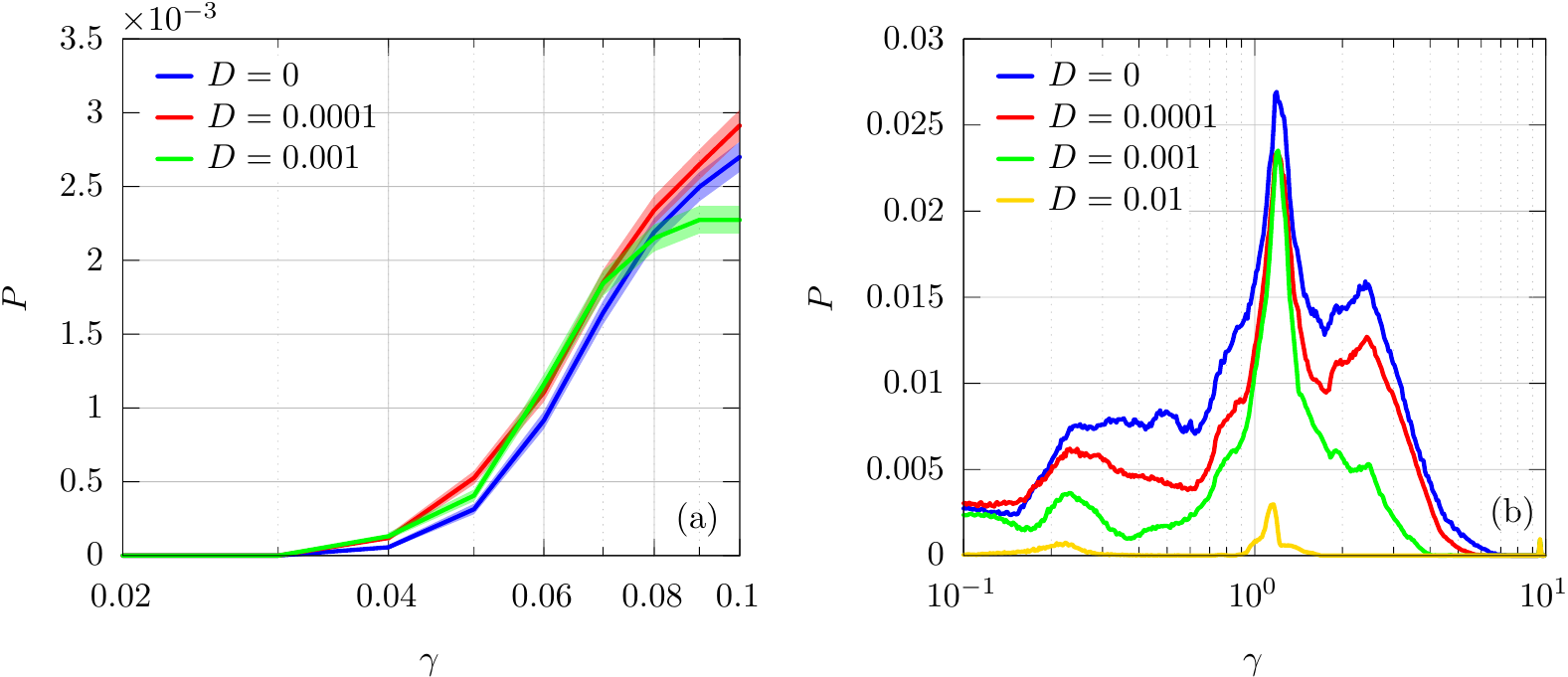}
	\caption{The empirical probability $P$ of observing the negative mobility effect as a function of the friction coefficient $\gamma$ for different values of temperature $D$. Panel (a) shows the probability $P$ in a weak damping regime. In both plots the static bias is fixed to $f = 0.1$. The shaded regions in panel (a) represent the 95\% confidence interval for the probability $P$ calculated using the exact Clopper-Pearson method \cite{Clopper1934}. In (b) it is smaller than the linewidth.}
	\label{fig:Pg}
\end{figure*}

\subsection{Quantity of interest}
We characterize the directed transport of the particle by its average velocity defined as
\begin{equation}
	\langle v \rangle = \lim\limits_{t\to\infty} \frac{1}{t} \int_0^t \mathrm{d}s\langle \dot{x}(s) \rangle,
\end{equation}
where $\langle \cdot \rangle$ indicates the average over all thermal noise realizations as well as different initial conditions for the particle position $x_0 = x(0)$, velocity $v_0 = v(0)$ and phase $\phi_0 = \phi(0)$ of the driving force $a \cos{(\omega t)}$. The latter is mandatory for the deterministic counterpart of the studied dynamics, i.e. for $D = 0$, when the system can be non--ergodic and consequently the result will be affected by the specific choice of the initial conditions.

Since the periodic potential $U(x)$ and external driving $a\cos{(\omega t)}$ are spatially and temporally symmetric while thermal fluctuations $\xi(t)$ vanish on average, the only perturbation that breaks the symmetry of the system is the static force $f$. 
Moreover, it easily follows that the average velocity $\langle v \rangle$ is an odd function of the bias $f$, i.e. $\langle v \rangle(-f) = -\langle v \rangle(f)$. 
For this reason from now on we will limit our consideration to non--negative values of the static force $f > 0$.
The average velocity $\langle v \rangle$ can be related with the static force $f$ via the mobility $\mu(f)$ \cite{Kostur2008} as
\begin{equation}
	\langle v \rangle = \mu(f) f.
\end{equation}
From the Green--Kubo relation it follows that in the small bias limit the mobility $\mu$ does not depend on the value of the static force $f$, i.e. $\lim_{f \to 0} \mu(f) = \mu_0$.
The intuitive case of $\mu_0 > 0$ corresponds to the Ohmic--like behaviour in which transport $\langle v \rangle > 0$ occurs in the direction of applied bias $f > 0$.
The counterintuitive situation \mbox{$\mu_0 < 0$} implies that transport $\langle v \rangle < 0$ emerges in the direction opposite to the static force $f > 0$.
This phenomenon is called the absolute negative mobility (ANM) \cite{Machura2007}.

In the linear response regime described above the average velocity $\langle v \rangle$ tends to zero only when the applied bias vanishes $f \to 0$.
For sufficiently large values of the static force the above statement is no longer true as the system is in the nonlinear response regime where in contrast to the previous case the mobility $\mu(f)$ depends on the applied bias. 
In such a case $\langle v \rangle$ can tend to zero even for $f \neq 0$. 
It means in particular that there might be some interval of the bias $f > 0$ away from $f = 0$ for which $\langle v \rangle < 0$ and consequently transport occurs in the negative direction.
Such a scenario is called the negative nonlinear mobility (NNM) \cite{Kostur2008}.

There are several conditions that the system needs to fulfill which allow to the emergence of the negative mobility \cite{Machura2007,Speer2007pre}.
Firstly, the superposition principle, which by definition holds in the linear systems, implies that the average velocity $\langle v \rangle > 0$ must have the same sign as the static force $f > 0$.
This means that only nonlinear systems can exhibit $\mu(f) < 0$.
In our case the nonlinearity is a consequence of the sinusoidal form of the potential $U(x)$.
Secondly, systems in thermal equilibrium obey the Le Chatelier--Braun's principle \cite{Landau}, which says that the system's response to a variation of its parameters causes a shift in the position of equilibrium that contradicts this change, i.e. the net velocity $\langle v \rangle > 0$ follows the direction of the constant bias $f > 0$.
For this reason the negative mobility $\mu(f) < 0$ requires a nonequilibrium state.
In our model the particle is driven out of the thermal equilibrium by the force $a\cos(\omega t)$.
Lastly, it is known that $\mu(f) < 0$ is forbidden when $m=0$ or $\gamma=0$, i.e. this effect cannot emerge in overdamped and Hamiltonian systems \cite{Speer2007pre}.

\section{Methods} \label{sec:methods}
Since Eq.~(\ref{eq:lang:gam}) is a nonlinear stochastic second order differential equation it cannot be solved analytically.
It is so also for the corresponding Fokker--Planck equation.
For this reason we studied our system by performing comprehensive numerical analysis.
We followed the methods used in Ref.~[\onlinecite{Wisniewski2022}] and implemented a weak second order predictor--corrector algorithm to solve Eq.~(\ref{eq:lang:gam}).
The timestep of the simulations was scaled as $h=0.01\times \mathsf{T}$, where $\mathsf{T}=2\pi/\omega$ is the fundamental periods of the driving force $a\cos(\omega t)$.
The average velocity $\langle v \rangle$ depending on the parameter regime was averaged over the ensemble of up to $10^6$ system trajectories, each starting with different initial conditions $x_0 \in [0, L=1]$, $v_0 \in [-2, 2]$ and with different initial phase $\phi_0 \in [0,2\pi]$ of the driving $a\cos(\omega t)$.
The time span of simulation depended on the parameter regime and ranged up to $10^9$ periods $\mathsf{T}$.
To increase the throughput we harvested the power of the Graphics Processing Unit (GPU) supercomputers. This innovative method \cite{Spiechowicz2015cpc} allowed us to analyze multiple trajectories in parallel and offered a speedup of a factor of the order $10^3$ as compared to the present day Central Processing Unit (CPU) approach.
\begin{figure}[t]
	\centering
	\includegraphics[width=\linewidth]{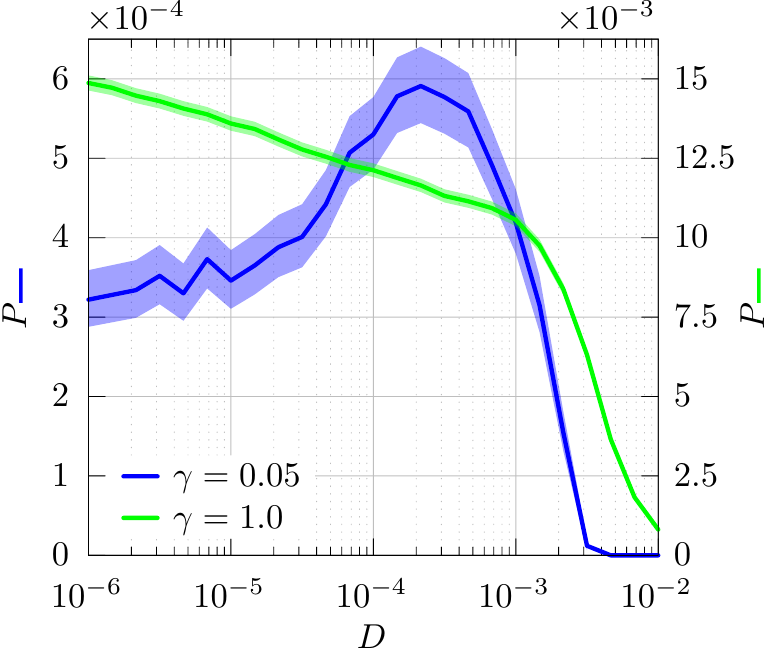}
	\caption{The empirical probability $P$ of observing the negative mobility effect as a function of temperature $D$ for $\gamma=0.05$ and $\gamma=1.0$. The static bias reads $f = 0.1$. The shaded regions represent the 95\% confidence interval for $P$ calculated using the Clopper--Pearson method.}
	\label{fig:PD}
\end{figure}

\subsection{Deterministic chaos detection}
Besides characterizing the directed transport, we were also interested in the emergence of deterministic chaos in our system. A common method for its detection is calculation of the Lyapunov exponent spectrum \cite{Ott2002}.
Let us consider an infinitesimal ellipsoid in the phase space of the system for the initial moment of time $t = t_0 = 0$. After the infinitesimal time $t$ we can approximate the length of the principal axes $l_i(t)$ of the ellipsoid as
\begin{equation} \label{eq:lyap:l}
	l_i(t) = l_i(0)e^{\lambda_i t},
\end{equation}
where $\lambda_i$ is the Lyapunov exponent corresponding to the given axis.
When $\lambda_i < 0$, the trajectories converge along the corresponding axis, whereas for $\lambda_i > 0$ they diverge.
\begin{figure*}[t]
	\centering
	\includegraphics[width=\linewidth]{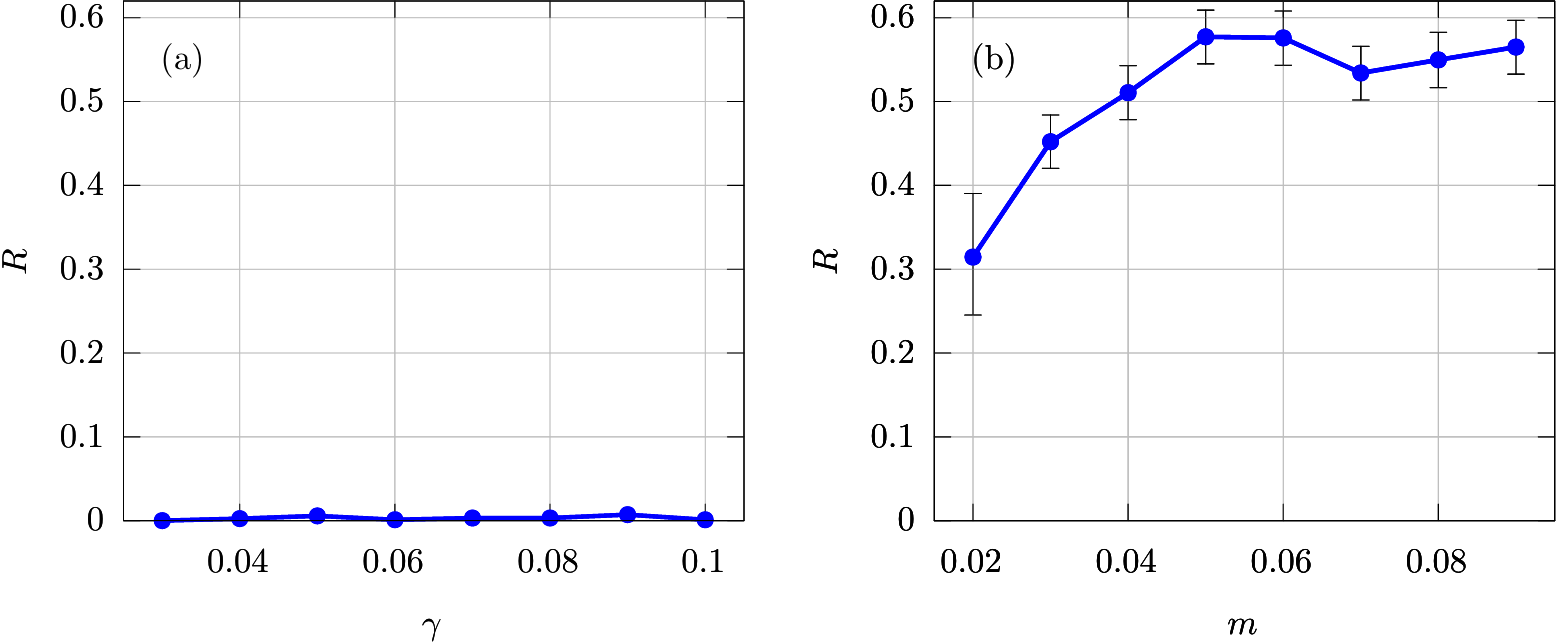}
	\caption{The fraction $R$ of parameter regimes exhibiting the deterministic negative mobility which displays chaotic behavior as a function of the dissipation $\gamma$ and the particle inertia $m$ in panel (a) and (b), respectively. Plot (a) corresponds to the weak damping regime $\gamma \ll 1$, whereas (b) to the limit of strong dissipation $m \ll 1$. Other parameters are: the static bias $f = 0.1$ and temperature $D = 0$. The error bars represent the 95\% confidence interval calculated according to the Clopper-Pearson method (in the left panel the error bars are smaller than the point size).}
	\label{fig:Rgm}
\end{figure*}

To analyze the Lyapunov spectrum for our system of interest, it is convenient to convert Eq.~ (\ref{eq:lang:gam}) to a set of autonomous ordinary differential equations
\begin{equation}
	\frac{\mathrm{d}\mathcal{X}(t)}{\mathrm{d}t} = \mathcal{F}(\mathcal{X}(t))
\end{equation}
where $\mathcal{X}(t)$ is the phase space vector and $\mathcal{F}(\mathcal{X}(t))$ specifies the vector field for the analyzed system of differential equations.
The deterministic variant of the studied model given by Eq.~(4) is described by a non--autonomous second--order differential equation with a time dependent force $a\cos(\omega t)$. Consequently, the phase space of the autonomous system is spanned by three variables corresponding to the position $x(t)$ of particle, its velocity $v(t)$ as well as the phase $\phi(t) = \omega t$ of the external driving force, i.e.
\begin{equation} \label{eq:psv}
	\mathcal{X}(t) = (x(t), v(t), \phi(t)).
\end{equation}
The vector field $\mathcal{F}(\mathcal{X}(t))$ then reads
\begin{equation}
	\mathcal{F}(\mathcal{X}(t)) = (v, -\gamma v - 2\pi\cos(2\pi x) + a\cos(\phi) + f, \omega).
\end{equation}
We can now consider an infinitesimal ellipsoid with volume $\mathcal{V}$ in the phase space of the system.
The evolution of its volume in time can be expressed as
\begin{equation}
	\frac{1}{\mathcal{V}} \frac{d \mathcal{V}}{d t} = \nabla \cdot \mathcal{F} = -\gamma,
\end{equation}
which yields
\begin{equation} \label{eq:lyap:V}
	\mathcal{V}(t) = \mathcal{V}(0)e^{-\gamma t}.
\end{equation}
It means that the phase space volume shrinks in time, as it should be for a dissipative system.
On the other hand, from Eq.~(\ref{eq:lyap:l}) it follows that
\begin{equation}
	\mathcal{V}(t) = \mathcal{V}(0) e^{(\lambda_x+\lambda_v+\lambda_\phi)t},
\end{equation}
which implies that $\lambda_x+\lambda_v+\lambda_\phi=-\gamma$.
$\lambda_\phi$ corresponds to the direction parallel to the system trajectory. Therefore it does not contribute to a change of the phase space volume. Setting $\lambda_\phi = 0$ yields
\begin{equation}
	\lambda_x + \lambda_v = -\gamma.
\end{equation}
To determine whether for a given parameter regime the system is chaotic it is sufficient to calculate only the maximum Lyapunov exponent $\lambda_{\mathrm{max}}=\mathrm{max}(\lambda_x, \lambda_v)$. When $\lambda_{\mathrm{max}} > 0$ it is chaotic whereas if $\lambda_{\mathrm{max}} \leq 0$ it is not.

Since the maximum Lyapunov exponent of our system cannot be calculated analytically, we estimated its value numerically.
One of the most common technique is to simulate the system trajectory until it reaches the asymptotic state, then shift it in the phase space by some small factor and observe how the distance between the original and shifted trajectories changes in time \cite{Wolf1985}.
Another method relies on reconstructing the attractor from a time series \cite{Rosenstein1993}.
These both algorithms are prone to numerical errors and therefore they may give inaccurate results.

Due to complexity of the analyzed system the first method did not always give reliable estimates of the maximum Lyapunov exponent.
Therefore we chose the latter one which does not depend on details of the studied model, but relies purely on the the system trajectory.
This method, however, sometimes also fails to give a correct estimate.
This usually happens when two points of the reconstructed trajectory are very close to each other and it leads to incorrectly high values of the estimated Lyapunov exponents for non--chaotic trajectories.
To eliminate this problem we made use of the fact that the autocorrelation function of a periodic trajectory repeatedly equals the unity, whereas for a chaotic time series it never reaches this value.
Putting together these two methods allowed us to detect the cases in which the numerical errors resulted in wrong estimates of the Lyapunov exponents.
\begin{figure*}[t]
	\centering
	\includegraphics[width=\linewidth]{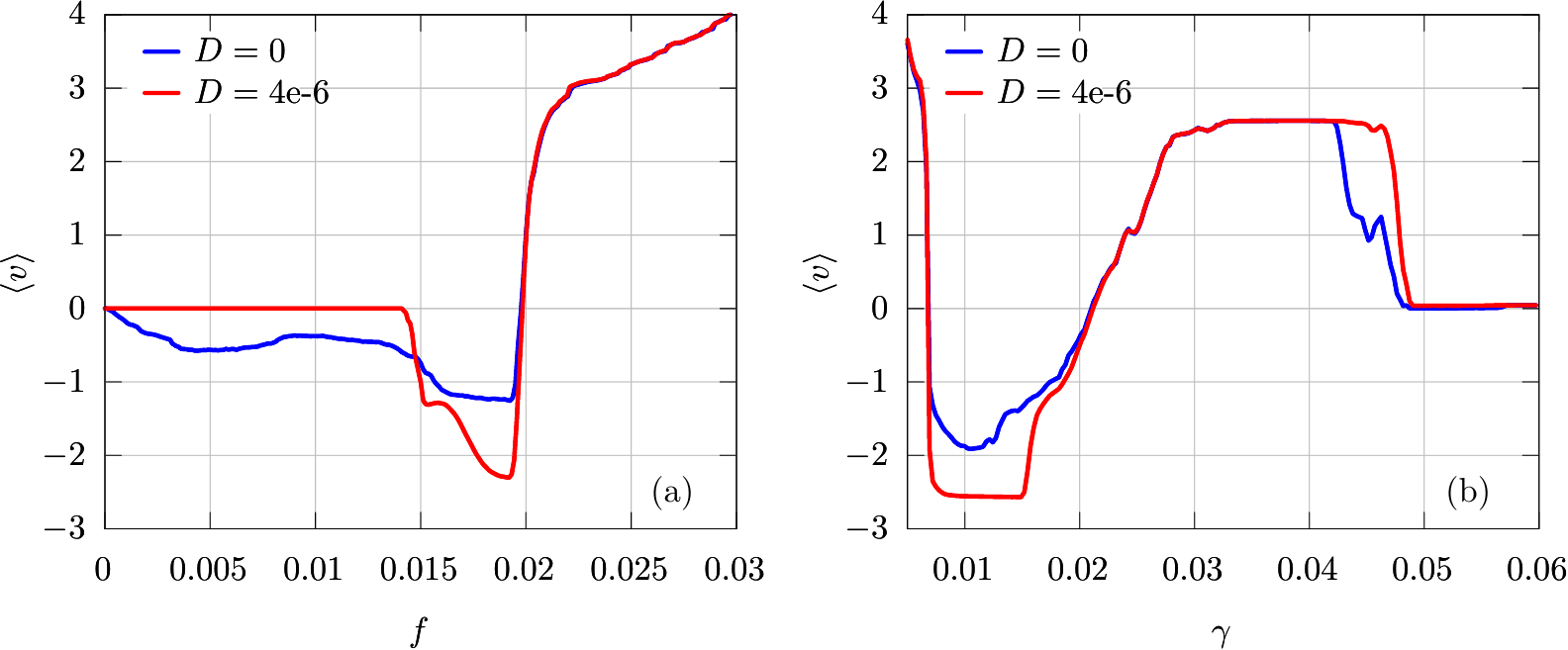}
	\caption{The average velocity $\langle v \rangle$ of the Brownian particle vs the static bias $f$ (panel (a)) and damping constant $\gamma$ (panel (b)) illustrating the emergence of negative mobility in the extremely weak dissipation regime. Parameters read $\gamma = 0.007$, $a = 4.818$, $\omega = 4.028$ and $f = 0.02$.}
	\label{fig:smallg}
\end{figure*}

\section{Results} \label{sec:results}
The studied system possesses a complex five dimensional parameter space $\{\gamma, a, \omega, f, D\}$ whose detailed and systematic exploration was not possible until very recently due to limited hardware capabilities as well as lack of innovative implementations of computational algorithms. In our latest work \cite{Wisniewski2022} we harvested GPU supercomputers to analyze the distribution of negative mobility effect in the parameter space of this system. We restricted our investigation to a subspace of the particle--environment coupling constant $\gamma \in [0.1,10]$. The choice of that range was dictated by the fact that the negative mobility neither can emerge in the dissipationless (Hamiltonian) regime, i.e. for $\gamma = 0$, nor it can be detected for the overdamped case, i.e. for $m = 0$ (see Ref.~[\onlinecite{Wisniewski2022}] for a discussion why it is not equivalent to $\gamma \to \infty$). In Ref.~[\onlinecite{Slapik2018}] this anomalous transport phenomenon was investigated in the limiting regime of strong damping $m \to 0$. In contrast, here we want to present several paradoxes of negative mobility for the remaining case of weak dissipation $\gamma \to 0$.

\subsection{Paradoxes of negative mobility in the weak dissipation regime}
The starting point of our analysis will be estimation of the empirical probability $P(\gamma)$ for the emergence of the negative mobility effect for a given value of dissipation $\gamma$. To calculate it we first fixed the static bias $f$ and temperature $D$. Next, we created a meshgrid $(a, \omega) \in [0,25] \times [0,20]$ of the external driving force amplitude $a$ and frequency $\omega$ with a resolution of 400 points per dimension. For 400 values of $\gamma \in [0.1,10]$ we counted the number of parameter regimes for which the negative mobility effect occurs. The ratio of this quantity to the total number of $(a, \omega)$ pairs yielded the empirical probability $P(\gamma)$ describing how often the particle mobility is negative for a given value of the friction coefficient $\gamma$. The calculations were performed for several values of $f$ but here we will restrict ourselves only to $f=0.1$. 

The result is shown in Fig.~\ref{fig:Pg} (b). The empirical probability $P(\gamma)$ for the emergence of negative mobility versus the dissipation constant is depicted there for different values of temperature $D$. The reader can observe that in most of the range of $\gamma$ temperature generally has destructive influence on the occurrence of the anomalous transport effect as this characteristic is maximal for the deterministic dynamics with $D = 0$. However, at the entrance of the weak dissipation regime $\gamma \approx 0.1$ one can see that the negative mobility emerges more frequently for the noisy $D = 0.0001$ rather than deterministic $D = 0$ system. It suggests that for this parameter region thermal noise plays a crucial role. This fact serves as the cornerstone of the analysis presented below.

\subsubsection{Thermal fluctuations play the leading role}
For the system described by Eq.~(\ref{eq:lang:gam}) intensity of thermal fluctuations $\xi(t)$ can be defined as $Q = \gamma D$. It implies that when temperature $D$ of the system is fixed, $Q$ decreases if the dissipation constant diminishes \mbox{$\gamma \to 0$}. Consequently, in the weak damping regime one can naively expect that the impact of thermal fluctuations on the system dynamics will be minimal. However, in Fig.~\ref{fig:Pg} (a) we show that actually the reverse is true, namely, in the weak dissipation limit $\gamma \ll 1$ thermal noise plays the leading role and its influence on the emergence of negative mobility effect is the most substantial. The plot of the empirical probability $P(\gamma)$ reveals that indeed the anomalous transport behavior occurs more frequently for the noisy $D \neq 0$ rather than deterministic $D = 0$ system. This fact suggest that the contribution of parameter regimes in which thermal noise induces the negative mobility is particularly significant. While in principle such mechanism is known in the literature \cite{Machura2007} to the best of our knowledge no one has discovered before that it manifests to the greatest extent in the least obvious limit of weak dissipation.

Moreover, when temperature $D$ grows the negative mobility phenomenon must finally vanish since at some point thermal fluctuations dominate the dynamics and other contributions necessary for the emergence of this effect become negligible. This claim can be seen in Fig.~\ref{fig:Pg} (b) already for $D=0.01$. It suggests that in the studied weak dissipation regime the impact of thermal fluctuations on the occurrence of negative mobility is additionally non--monotonic. We confirm this remark in Fig.~\ref{fig:PD} where we show the empirical probability $P(D)$ for the emergence of anomalous transport behavior as a function of temperature $D$ in the near--Hamiltonian regime $\gamma = 0.05$ and for the fixed static bias $f = 0.1$. The reader can observe there the non--monotonic dependence. Conversely, a similar curve for $\gamma=1.0$, also depicted in Fig.~\ref{fig:PD}, is strictly decreasing with $D$. Although Fig.~\ref{fig:PD} presents only empirical estimate of the real values of the probability $P(D)$, the substantial difference between the curves for $\gamma=0.05$ and $\gamma=1.0$ supports very well our previous claim that thermal fluctuations play the leading role in the emergence of negative mobility in the weak dissipation regime.

\subsubsection{Deterministic negative mobility is non--chaotic}
As we just illustrated, in the weak damping regime there exist a significant number of parameter regimes for which the negative mobility is induced by thermal fluctuations. However, what is the mechanism of this effect for a set of parameters where it is rooted in the deterministic dynamics is a problem that remains to be solved. In such a case its origin may lay either in a regular attractor transporting the particle $\langle v \rangle < 0$ in the direction opposite to the acting static bias $f > 0$ \cite{Slapik2018} or in a complex chaotic dynamics \cite{Machura2007,Speer2007epl}. 

The main consequence of the Poincar{\'e}-Bendixson theorem \cite{Ott2002} is that continuous dynamical systems whose phase space has less than three dimensions cannot exhibit chaotic behavior. It means that chaos is ruled out for the system described by Eq.~(\ref{eq:lang:gam}) within the overdamped approximation $m = 0$, see also Eq.~(\ref{eq:psv}). On the other hand, chaotic behavior is characteristic feature of many systems in the opposite limit of dissipationless regime with $\gamma = 0$. These two facts implies that naively the deterministic chaos should be detected more frequently for the weak damping $\gamma \ll 1$ rather than when the dissipation is strong $m \ll 1$.

In Fig.~\ref{fig:Rgm} (a) we present the fraction $R$ of parameter regimes exhibiting the deterministic negative mobility which displays chaotic behavior as a function of the dissipation strength $\gamma$. The classification was done using the analysis of the maximal Lyapunov exponent supported by the autocorrelation function of the particle velocity trajectory. Paradoxically, it turns out that the intuitive prediction is wrong as in the weak dissipation regime $\gamma \ll 1$ presented there chaos can be barely observed.
It means that the deterministic negative mobility is rooted solely in regular, non--chaotic behavior when approaching the Hamiltonian regime. Moreover, in the panel (b) of the same figure we present the fraction $R$ for the opposite limit of the strongly damped $m \ll 1$ particle whose dynamics is described by Eq.~(\ref{eq:lang:mass}). The result is again counterintuitive as in more or less half of the studied parameter sets the system exhibits chaotic behavior whereas in the overdamped regime $m = 0$ it ceases to exist. Summarizing, we exemplified a scenario in which physical properties in the overdamped or Hamiltonian regimes cannot serve as a naive guide for features in strongly or weakly damped situation, respectively.
\begin{figure}[t]
	\centering
	\includegraphics[width=\linewidth]{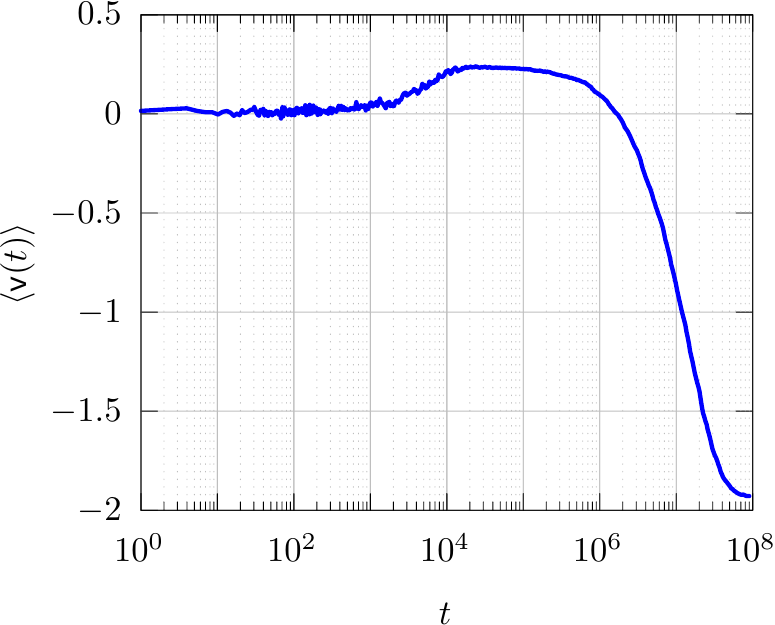}
	\caption{Time evolution of the velocity $\langle \mathsf{v}(t) \rangle$ of the particle averaged over the period of external driving $a\cos(\omega t)$ for $f=0.004$, $D=2.3\times10^{-6}$, $\gamma=0.08$, $a=3.3$ and $\omega=4.04$.}
	\label{fig:traj_longrelax}
\end{figure}

\subsubsection{Hamiltonian vs weak dissipation regime}
In this spirit we note that the negative mobility does note emerge in the Hamiltonian regime. Neglecting the dissipative term $-\gamma v(t)$ in Eq.~(\ref{eq:lang:gam}) means $\gamma \to 0$ while keeping thermal noise intensity $Q = \gamma D$ fixed. It implies $D \to \infty$, i.e. the particle is coupled to an infinitely hot thermal bath. Then the impact of periodic potential $U(x)$ in principle becomes negligible. By virtue of the superposition principle valid for linear systems this fact rules out the negative mobility. We now pose an important question what is the lower limit of dissipation $\gamma$ below which this effect ceases to exist.
\begin{figure*}[t]
	\centering
	\includegraphics[width=\linewidth]{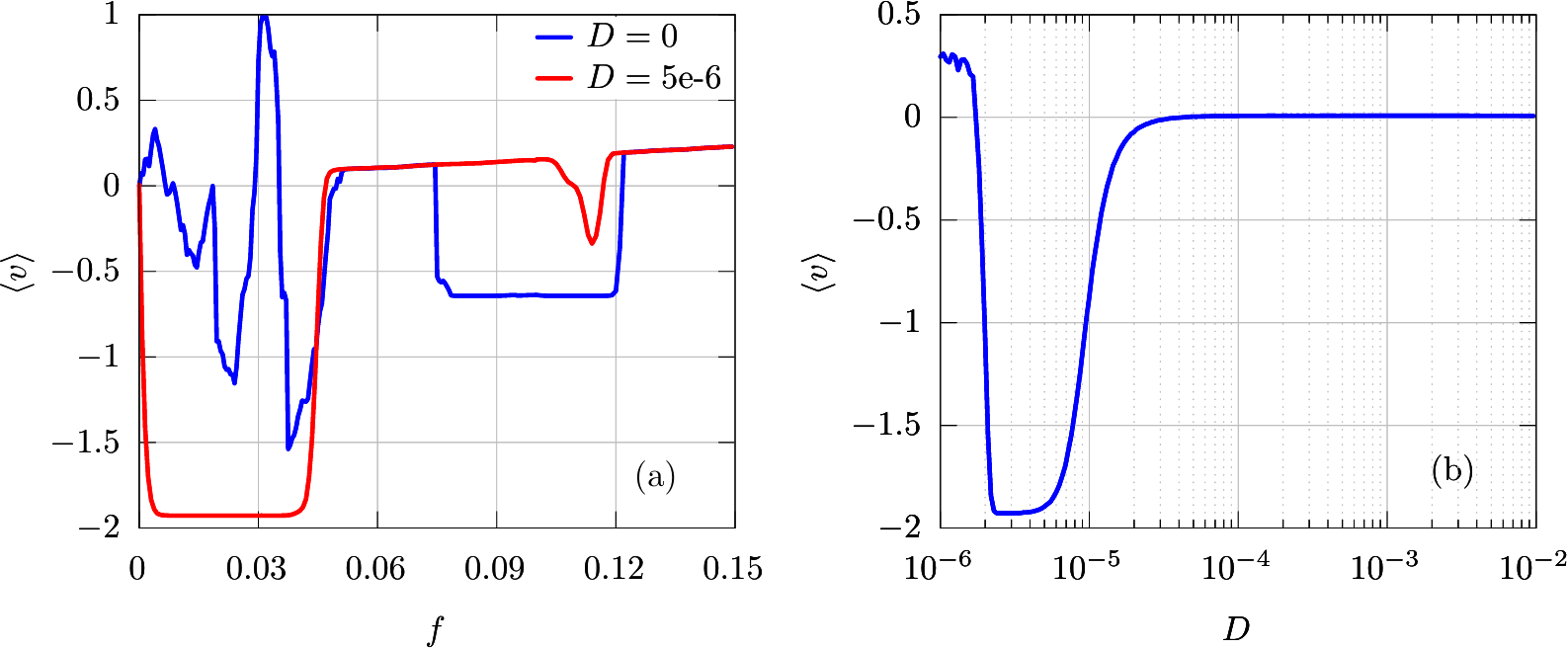}
	\caption{The average velocity $\langle v \rangle$ of the Brownian particle versus the static bias $f$ (panel (a)) and temperature $D$ (panel (b)). The parameters values are $\gamma=0.08$, $a=3.3$ and $\omega=4.04$. In (b) the external force reads $f = 0.004$.}
	\label{fig:induced_ANM}
\end{figure*}

In Fig.~\ref{fig:smallg} we illustrate the parameter set for which the studied anomalous transport behavior emerges in the extremely weak dissipation regime $\gamma < 0.01$. In panel (a) the average velocity $\langle v \rangle$ of the Brownian particle is depicted as a function of the static bias $f$, whereas in (b) the same quantity is presented versus the damping constant $\gamma$. One can note that the negative mobility is rooted in the deterministic dynamics as it survives even in zero temperature $D = 0$ limit. In the latter case this effect is detected in the linear regime as the external force $f$ grows. The impact of weak thermal fluctuations is twofold: they reduce the range of bias $f$ where the negative mobility emerges but at the same time they enhance the anomalous response in the nonlinear regime. Applying sufficiently high temperature will eventually terminate the negative mobility. 

\subsubsection{Weak ergodicity breaking}
The deterministic dynamics corresponding to Eq.~(\ref{eq:lang:gam}) with $D = 0$ exhibits extremely rich and complex behavior. Depending on the parameter regime, periodic, quasiperiodic and chaotic motion can be observed \cite{Kautz1996}. Typically it is accompanied by the strong ergodicity breaking \cite{Spiechowicz2016scirep} that manifests itself as a partition of the phase space into invariant and mutually inaccessible attractors. In the latter case the initial conditions are never forgotten. On the other hand, at non--zero temperatures the system is in principle ergodic as thermal fluctuations of intensity $Q = \gamma D$ activate stochastic escape events connecting coexisting deterministic disjoint attractors \cite{Spiechowicz2022entropy}. However, the time after the phase space of the system is fully sampled depends on noise intensity $Q$. If $Q \to 0$, which is indeed the case for the weak dissipation regime $\gamma \to 0$ with fixed temperature $D$, it may be extremely long and go to infinity. Such a situation is often termed as the weak ergodicity breaking \cite{Spiechowicz2016scirep} and can be identified with an unusually slow relaxation of the system towards its stationary state. It can be characterized by the Deborah number
\begin{equation}
	\mbox{De} = \frac{\tau}{\mathcal{T}}
\end{equation}
which is the ratio of a relaxation time $\tau$ of a given observable and the time of observation. For the weak ergodicity breaking it diverges $\mbox{De} \to \infty$. It can happen because of (i) $\mathcal{T}$ is too short but also, more interestingly, (ii) $\tau$ is extremely long.
\begin{figure*}[t]
	\centering
	\includegraphics[width=\linewidth]{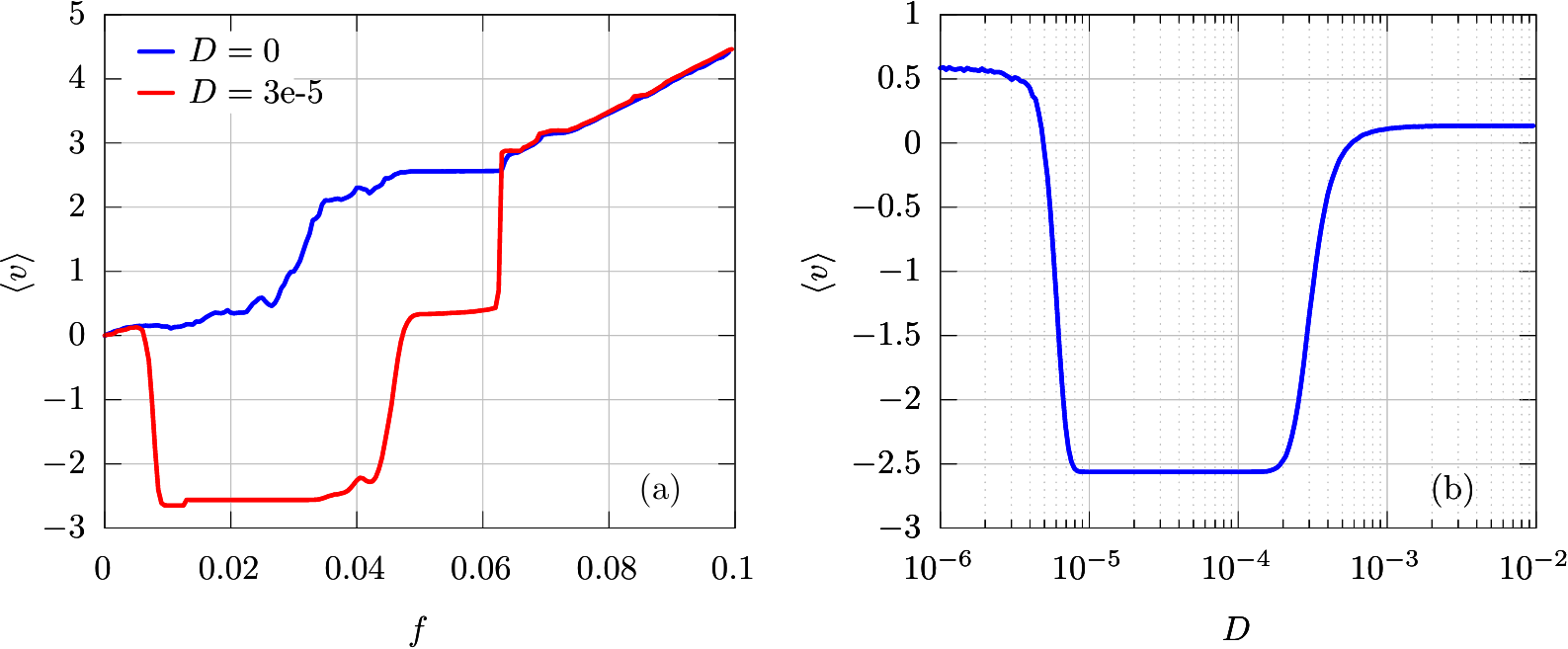}
	\caption{The directed velocity $\langle v \rangle$ of the Brownian particle versus the static bias $f$ and temperature $D$ in panel (a) and (b), respectively. The parameters values are $\gamma=0.022$, $a=4.818$ and $\omega=4.028$. In (b) the external force reads $f = 0.025$.}
	\label{fig:induced_NNM}
\end{figure*}

In Fig.~\ref{fig:traj_longrelax} we present the latter case which is typical for the weak dissipation regime. As an observable of interest we pick the velocity
\begin{equation}
	\langle \mathsf{v}(t) \rangle = \frac{1}{\mathsf{T}} \int_t^{t+\mathsf{T}} ds \langle \dot{x}(s) \rangle
\end{equation}
averaged over the period $\mathsf{T} = 2\pi/\omega$ of external driving $a \cos{(\omega t)}$ which in contrast to the instantaneous velocity $v(t)$ possesses the stationary state \cite{Jung1993}. The relation between the studied directed velocity $\langle v \rangle$ and $\mathsf{v}(t)$ can be expressed as
\begin{equation}
	\langle v \rangle = \lim_{t \to \infty} \langle \mathsf{v}(t) \rangle.
\end{equation}
Time evolution of $\langle \mathsf{v}(t) \rangle$ depicted in Fig.~\ref{fig:traj_longrelax} reveals that its relaxation towards the stationary state is extremely slow as even after $10^6$ dimensionless time units the particle velocity is positive while eventually the negative mobility emerges. Therefore this anomalous transport phenomenon in the weak dissipation regime is often affected by the weak ergodicity breaking. It makes it very challenging to investigate since special carefulness is needed to not confuse the negative mobility with transiently negative velocity or overlook this effect because the stationary state has not yet been reached. For this reason any quantities of interest that require the analysis of a large number of parameter regimes -- such as e.g. the empirical probability $P$ for the emergence of negative mobility -- for which the time span of evolution is always a compromise should be treated as a guide towards physical reality rather than taken as granted without ``errors''.

\subsection{Constructive influence of thermal fluctuations}
In previous subsection we discussed several paradoxes characteristic for the negative mobility effect in the weak dissipation regime. In particular, we demonstrated that in contradiction to common intuition in this limiting case thermal fluctuations play the leading role in emergence of this phenomenon, see Fig.~\ref{fig:PD}. Here we exemplify two possible scenarios of constructive influence of thermal noise which contribute to such behavior.

\subsubsection{Thermal noise induced absolute negative mobility}

The first example is induction of the absolute negative mobility by thermal fluctuations.
In this case in the small bias limit $f\to0$ the mobility $\mu_0 = \lim_{f\to0}\mu(f)$ is positive for $D=0$ and negative for $D\neq0$.
This effect is presented in Fig.~\ref{fig:induced_ANM}.
Panel (a) shows the dependance of the average velocity $\langle v \rangle$ on the static force $f$.
In the deterministic system $D=0$ the mobility $\mu_0$ is positive for small values of bias $f$.
Weak fluctuations $D=5\times10^{-6}$ completely change the response of the system leading to $\mu_0<0$.
The plot $\langle v \rangle(D)$ presented in panel (b) reveals that this effect can be observed only for small values of the thermal noise intensity $D$.
Higher temperatures lead to $\mu_0=0$ and the transport in small bias regime ceases to exist.

\subsubsection{Thermal noise induced nonlinear negative mobility}
Another manifestation of the constructive influence of fluctuations is thermal noise induced nonlinear negative mobility. In Fig.~\ref{fig:induced_NNM} we report this effect. In panel (a) the directed velocity $\langle v \rangle$ of the Brownian particle is depicted as a function of the static bias $f$. For the deterministic system $D = 0$ it is non--negative for all values of $f$ whereas for weak fluctuations $D = 3 \times 10^{-5}$ the average velocity $\langle v \rangle$ initially increases as $f$ grows to later become negative thus portraying the nonlinear negative mobility. This finding is confirmed in panel (b) where the same quantity is depicted versus temperature $D$. Thermal noise induces the nonlinear negative mobility. While the counterpart of this effect for absolute negative mobility is known in literature \cite{Machura2007} to the best of author's knowledge in the nonlinear response regime it has not yet been reported for thermal equilibrium fluctuations.

\section{Conclusions} \label{sec:conclusions}

In this work we considered the paradigmatic model of nonequilibruim statistical physics, namely, a Brownian particle dwelling in a periodic potential and driven by both a constant and time--periodic force. We focused our discussion on the weak dissipation regime in which the friction constant is very small in comparison to the particle inertia. We revealed several paradoxes of the negative mobility phenomenon in which, counterintuitively, the net movement of the particle is opposite to the direction of the static force acting on it.

Firstly, the weak dissipation regime via the fluctuation--dissipation theorem implies the delicate interaction of the particle with its environment. Then, one naively expects that the influence of thermal fluctuations will be negligible. On the contrary, we demonstrated that in the weak damping regime thermal noise has the greatest impact on the emergence of the negative mobility effect. Thermal fluctuations usually enhance or induce this phenomenon, which happens much less frequently when the dissipation is stronger.

Secondly, we reported a counterintuitive relation between the deterministic negative mobility and chaos. The latter does not occur for the considered system in the overdamped regime while it is very frequent for the Hamiltonian limit of dissipationless dynamics. This fact may suggest that the deterministic negative mobility should be assisted by chaotic behavior more frequently for the weak dissipation regime rather than when the damping is strong. However, the reverse is actually true, we found that the coexistence of chaos and negative mobility is much more common when approaching the overdamped limit and it ceases to exist when the friction is small. This means that although chaos is often promoted by weak dissipation, typically it does not coincide with the negative mobility effect.

Thirdly, we exemplify that the negative mobility effect in the weak damping regime is often affected by the weak ergodicity breaking. It makes it very challenging to investigate since the particle velocity exhibits extremely slow relaxation and special carefulness is needed to not confuse the negative mobility with transiently negative velocity or overlook this effect because the stationary state has not yet been reached.

Fourthly, we confirmed that the negative mobility phenomenon does not emerge for the Hamiltonian (dissipationless) limit of the dynamics but surprisingly it can be detected even for as weak dissipation as $\gamma = 0.007$ while at the same time the dimensionless inertia reads $m = 1$.

Finally, we illustrated two instances of constructive influence of thermal fluctuations on the considered anomalous transport behavior. In particular, for the first time we show that thermal noise can induce the negative mobility in the nonlinear response regime, i.e. for specific values of the parameters of the system the net velocity of the particle is opposite to the biasing force only in the non--deterministic case.

Summarizing, this work shows that even the already highly counterintuitive phenomenon of negative mobility can display paradoxical features. Since the studied system is the paradigmatic model of nonequilibrium statistical physics, our theoretical findings can be verified experimentally in a number of accessible systems that embody it, see the last paragraph of Sec.~\ref{sec:model}.

\section*{Acknowledgment}
This work has been supported by the Grant NCN No.~2022/45/B/ST3/02619 (J.~S.)

\bibliography{references}

\end{document}